\documentclass[10pt]{IEEEtran}
\IEEEoverridecommandlockouts

\usepackage{graphics,graphicx} 
\usepackage[]{algorithm2e}

\usepackage{amsmath}

\usepackage{amssymb}
\usepackage{tikz}
\usepackage{amsthm}
\usepackage{comment}
\usepackage[export]{adjustbox}
\usepackage{mathcomSTEP}
\usepackage{subcaption}
\usepackage[utf8]{inputenc}
\usepackage{bm}

\usepackage{tikz}
\usetikzlibrary{positioning}
\usetikzlibrary{arrows,fit}

\usepackage{epstopdf}
\usepackage{pgfplots,setspace}

\allowdisplaybreaks
\onecolumn
\newtheorem{prop}[thm]{Proposition}

\newtheorem{definition}{Definition}

\tikzstyle{int}=[draw, fill=blue!10, minimum height = .5 cm, minimum width=1cm,thick ]
\tikzstyle{sum}=[circle, fill=blue!10, draw=black,line width=1pt,minimum size = 0.3cm, thick ]

\title{
Optimal Rate Allocation in \\
 Mismatched Multiterminal Source Coding
}

\author{

\IEEEauthorblockN{Ruiyang Song\IEEEauthorrefmark{1}, Stefano Rini\IEEEauthorrefmark{2},  Alon Kipnis\IEEEauthorrefmark{3}, and  Andrea J. Goldsmith\IEEEauthorrefmark{3} \\}

\IEEEauthorblockA{%
\IEEEauthorrefmark{1}
Tsinghua University, Beijing, China
}

\IEEEauthorblockA{%
\IEEEauthorrefmark{2}
National Chiao-Tung University, Hsinchu, Taiwan
}

\IEEEauthorblockA{%
\IEEEauthorrefmark{3}
Stanford University, CA, USA \\
Emails: \texttt{songry12@mails.tsinghua.edu.cn, stefano@nctu.edu.tw, \\ kipnisal, andrea@wsl.stanford.edu}
}
}

\begin{document}
%
\maketitle
%
\begin{abstract}
We consider a multiterminal source coding problem in which a source is estimated at a central processing unit from lossy-compressed remote observations.
Each lossy-encoded observation is produced by a remote sensor which obtains a noisy version of the source and compresses this observation 
minimizing a local distortion measure which depends only on the marginal distribution of its observation.
%
The central node, on the other hand, has knowledge of the joint distribution of the source and all the observations and  produces the source estimate which minimizes a different distortion measure between the source and its reconstruction.
In this correspondence, we investigate the problem of optimally choosing the rate of each lossy-compressed remote estimate so as to minimize the distortion at the central processing unit subject to a bound on the overall communication rate between the remote sensors and the central unit.
We focus, in particular, on two models of practical relevance:  the case of a Gaussian source observed in additive Gaussian noise and reconstructed under quadratic distortion, and the case of a binary source observed in bit-flipping noise and reconstructed under Hamming distortion.
In both scenarios we show that there exist regimes under which having more remote encoders does reduce the source distortion: in other words, having fewer, high-quality remote estimates provides a smaller distortion than having more, lower-quality estimates.
%
%
 \end{abstract}

\section{Introduction}
The Distortion-Rate Function (DRF) characterizes the minimum code-rate required for encoding an information source as to ensure its recovery to be within a prescribed average distortion.
A natural extension of this setup is the case in which the encoder can only observe the source through noisy observations: this scenario is referred to as the \emph{remote} or \emph{indirect} source coding problem \cite[Sec. 3.5]{berger1971rate}.
%
%
The setting in which multiple remote encoders aid the reconstruction of a remote source at a central decoder corresponds to the Chief Executive Officer (CEO) problem \cite{berger1996ceo}.
%
Unfortunately, the CEO problem is intrinsically arduous to solve. Hence, despite its relevance in many sensing applications, only a few results for this model are available in the literature \cite{prabhakaran2004rate, 6651793}.
%

Given the difficulty in deriving the optimal rate-distortion trade-off in the CEO problem, one naturally wonders if there exists a more tractable model which addresses the multiterminal remote source coding problem.
Indeed, the mismatched multiterminal source coding problem in \cite{kipnis2016mismatch} is a step in this direction as it considers the case in which the distortion criterion at the remote encoders does not depend on the source distribution or the decoder distortion measure.
This is a different setup than the CEO problem, but a judicious choice of the distortion measure for the remote encoders produces similar performances
 \cite{kipnis2016mismatch}.
Since the codebooks employed by the remote encoders are not necessarily optimal in minimizing the distortion criterion at the central node, this problem is referred to as the \emph{mismatched} multiterminal source coding.
%
In the mismatched multiterminal source coding problem, the encoding operations at the remote encoders are chosen among a family of possible encoding rules and the reconstruction operation is determined by the optimal source estimate given the remote lossy-compressed samples.
Specifically, it is assumed that these encoding rules are chosen to operate at the optimal rate-distortion point asymptotically as the block length goes to infinity with respect to the local distortion measure at each encoder.
Using properties of distributions that achieve the DRF, it was shown in \cite{kipnis2016mismatch} that the minimal distortion in reconstructing the source approaches a single letter expression as the blocklength of the encoders goes to infinity.
The  optimal distortion-rate trade-off for this scenario is described by the mismatched Distortion Rate Function (mDRF).
The availability of such a single letter expression for the mDRF allows us to evaluate the distortion-rate trade-off in closed form and moreover, as explored in this paper, to determine the optimal rate-allocation strategy given a prescribed target distortion.

%
The optimal rate-allocation in the mismatched multiterminal problem is of great practical relevance in distributed sensing networks, in which the remote sensors are either unaware of the existence of the underlying source or lack the flexibility and the resources to adapt to multiple sensing scenarios.
Under this setting, the sensors may have very different quality of measurements, forcing the central processor to allocate different communication resources and associated data rates to different sensors.

The remainder of this paper is organized as follows: in Sec \ref{sec:Problem Formulation} we introduce the problem formulation, while in Sec. \ref{sec:related results} we present relevant results available in the literature.
The result for the Gaussian source is analyzed in Sec. \ref{sec:Gaussian source+Gaussian noise+quadratic distortion};
the case for the binary source is analyzed in Sec. \ref{sec:Binary source+bitflipping noise+Hamming distortion}.
Finally, Sec. \ref{sec:Conclusions} concludes the paper.
%

\section{Problem Formulation}
\label{sec:Problem Formulation}

\begin{figure}
\centering
\begin{tikzpicture}[node distance=2cm,auto,>=latex]
  \node at (-5,0) (source) {$X^n$} ;
  \node [int, right of = source,node distance = 1.6cm](pxy2){$\mathrm P_{Y_2|X}^n$};
  \node [int, above of = pxy2,node distance = 1.2 cm](pxy1){$\mathrm P_{Y_1|X}^n$};
  \node [below of = pxy2,node distance = .8 cm](dots){$\vdots$};
  \node [int, below of = pxy2,node distance = 2 cm](pxyM){$\mathrm P_{Y_L|X}^n$};
\node [int,right of = pxy2, node distance = 2.2cm](enc2){$\mathrm{Enc. \ 2}$};
\node [int, above of = enc2,node distance = 1.2 cm](enc1){$\mathrm{Enc. \ 1}$};
\node [below of = enc2,node distance = 1 cm](dots){$\vdots$};
\node [int, below of = enc2,node distance = 2 cm](encM){$\mathrm{Enc. \  L}$};

%
\draw[->,line width=1pt] (pxy1) -- node[above,yshift =.25 cm] {$Y_1^n$}(enc1);
\draw[->,line width=1pt] (pxy2) -- node[above,yshift =.25 cm] {$Y_2^n$}(enc2);
\draw[->,line width=1pt] (pxyM) -- node[above,yshift =.25 cm] {$Y_L^n$}(encM);
\node [int] (dec) [right of=enc2, node distance = 2.3 cm] {$\mathrm{Dec}$};
\node [right] (dest) [right of=dec]{$\Xh^n(\mathbf W)$};
\draw[->,line width=1pt] (source) |- (pxy1);
\draw[->,line width=1pt] (source) |- (pxy2);
\draw[->,line width=1pt] (source) |- (pxyM);
\draw[->,line width=1pt] (enc1) node[above, xshift =1.3cm,yshift =.25 cm] {$W_1$} -|  (dec) ;
\draw[->,line width=1pt] (enc2) node[above, xshift =1.2cm,yshift =.25 cm] {$W_2$} -- (dec) ;
\draw[->,line width=1pt] (encM) node[above, xshift =1.3cm,yshift =.25 cm] {$W_L$} -| (dec) ;
\draw[->,line width=1pt] (dec) -- (dest);

\end{tikzpicture}
\caption{
The multiterminal mismatched distortion rate problem.
 }
\label{fig:distributed_scheme}
\vspace{-.5 cm}
\end{figure}
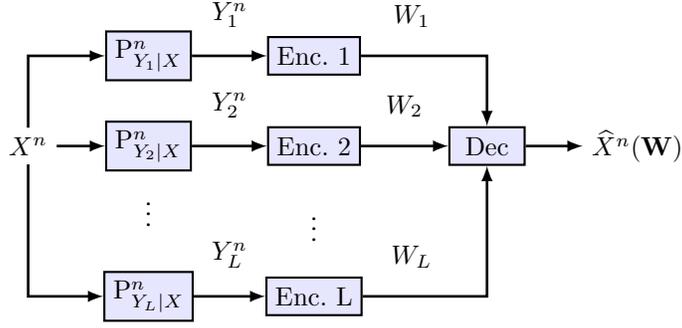

Our starting point is the source coding setting of Fig.~\ref{fig:distributed_scheme}: the source  sequence $X^n =(X_1,\ldots,X_n)$ is obtained through $n$ iid draws from the distribution $P_X(x)$. This source is observed at $L$ remote encoders through $n$ uses of the  memoryless channel
$P_{Y_l|X}$, ie:
\ea{
P_{Y_l^n|X^n}(y_l^n|x^n)=\prod_{i=1}^n P_{Y_l|X}(y_{l,i}|x_i), \quad l\in \{1 ,\ldots,L \}.
}
Given the observation $Y_l^{(n)}$, each remote encoder $\Ecal_l^{(n)}$ produces an index $W_l \in \{1 \ldots 2^{\lfloor n R_l \rfloor} \} $. The decoder receives $\Wv = (W_1,\ldots,W_L)$ and produces the sequence $\widehat{X}^n(\Wv)$. We consider the distortion between the reconstruction $\widehat{X}^n$ and the original source realization defined as
\ea{\label{eq:distortion_def}
D^{n} \leq D(X^n,\Xh^n)  \triangleq \f 1 n \sum_{i=1}^n \Ebb \left [d(X_i,\Xh_i)\right],
}
for some positive defined per-letter distortion $d(x,\xh)$.
Given $\Wv$, the decoder produces a  reconstruction sequence $\widehat{X}^n$ minimizing \eqref{eq:distortion_def}.

An achievable distortion and the optimal distortion-rate trade-off are defined as follow:
for a family of encoders of block-length $n$:
$
\Ecal^{(n)} \triangleq \Ecal_1^{(n)}  \times\cdots\times \Ecal_L^{(n)},
$
and an $n$-sequence distortion measure $D^n$,  we introduce the following definition:
\begin{definition}
A distortion $D$ is \emph{achievable} for encoders $\Ecal^{(n)}$ if there exist $L$ encoders $\left(g_1,\ldots,g_L\right) \in \Ecal^{(n)}$ such that $D^n\leq D$.
\end{definition}
The mDRF arises when, for block-length $n$, the encoders $\Ecal^{(n)}$ are chosen as in the classical source coding problem with source $Y_l^n$ and distortion criterion
\ea{
D_l^n (Y_l^n,\Yh_l^n) \triangleq  \f 1 n \sum_{i=1}^n  \Ebb \lsb  d_l(y_i,\yh_i) \rsb,
\label{eq:distortion y}
}
for some per-letter distortion $d_l(y,\yh)$. More precisely,
we require that the family of encoders $\Ecal^{(n)}$ is such that the optimal distortion-rate performance with respect to the sequence $Y_l^n$ and distortion $d_l$ is asymptotically achieved for all $l=1,\ldots,L$. Namely, {$D_l^n(Y_l,\Yh_l) \to D_l(R_l)$} while $I(Y_l;\Yh_l) \goes R_l$ where $D_l(R_l)$ is the DRF for the problem of reconstructing $Y$ under per-letter distortion $d_l$ at rate $R_l$.
Under these conditions, it is possible to obtain the following characterization of the distortion in estimating the source $X^n$ under this family of encoders, denoted by $D(\Ecal(R_1 \ldots R_L))$ \cite[Th. 4.1]{kipnis2016mismatch}:
\begin{equation}
\label{eq:main_thm}
D(\Ecal(R_1 \ldots R_L)) \triangleq  {\inf D(X,\Xh )},
\end{equation}
where the infimum is over all possible random mappings $(\Yh_1 \ldots \Yh_L) \rightarrow \widehat{X}$, where $p({\widehat{y}_1|y_1}), \ldots ,p({\widehat{y}_L|y_L}) $ are $L$ conditional distributions for which $I(Y_l;\Yh_l)= R_l$, and $D_l(Y_l,\Yh_l) = D_l(R_l)$.  We will refer to the function $D(\Ecal(R_1 \ldots R_L))$ as the \emph{mismatched multiterminal} distortion-rate function (mDRF).

Note that, unlike the traditional source coding definitions,  the mDRF is not defined as an optimization of encoding-decoding schemes with respect to the source $X$.
 It is only  defined as the attainable minimal distortion once the $L$, $n$-block-length encoders are determined.
 The only requirement for these encoders is that, as $n$ goes to infinity, they achieve the  rate-distortion function with respect to their local distortion measure and input sequence.
The problem considered in this correspondence is as follows: given a total communication rate-budget of $R$ bits, we ask what is the minimal value of $D(\Ecal(R_1 \ldots R_L))$ subject to the constraint
\ea{
\sum_{l=1}^L R_l \leq R,
\label{eq:sum rate constraint}
}
for some $R \in \Rbb^+$.

We focus, in particular, on two models:

\noindent
$\bullet$
{\bf Gaussian source in Gaussian noise and quadratic distortion:}
the source  $X\sim \Ncal(0,1)$ and the $l$th observation is
\begin{equation}
Y_l = \sqrt{\ga_l} X + Z_l, \quad l\in [1,\ldots,L],
\label{eq:gaussian+quadratic}
\end{equation}
for  $Z_l \sim \Ncal(0,1)$ and iid with $0 \leq \ga_L \leq \ldots \leq \ga_1$, while $d_l(u,\uh)=d(u,\uh)=(u-\uh)^2$.

\noindent
$\bullet$
{\bf Binary source in bit-flipping noise and Hamming distortion:}
the iid source follows Bern$(1/2)$ and the $l$th observation is
\ea{
Y_l = X \oplus Z_l, \quad l\in [1,\ldots,L],
\label{eq:binary +bitflip}
}
for $Z_l \sim {\rm Bern}(p_l)$ with $0 \leq p_1 \leq  \ldots \leq  p_L$ while $d_l(u,\uh)=d(u,\uh)= u \oplus \uh$.

\noindent

{\bf Notations:} In the following sections, we denote $\xo=1-x$, and $x \star y =x \yo+\xo y $. Moreover  $h$ denotes the binary entropy, i.e. $h(x)=x \log x +\xo \log \xo$. With  $h^{-1}(x)$ for $0\leq x \leq 1/2$ we indicate the inverse  of the binary entropy function, which is well defined in this interval.

\section{Related Results}
\label{sec:related results}

The source coding setting in which a single encoder only has partial information about the source is usually studied in the setting of universal source coding \cite[Ch. 11.3]{cover2012elements}. Another approach is to encode with respect to a min-max penalty over a family of source distributions \cite{dembo2003minimax}. Mismatched encoding can also arise in the case in which the codebook at the decoder is fixed but its encoding codebook is not \cite{lapidoth1997role}.

The most related multiterminal setting to this work is the CEO problem introduced in \cite{berger1996ceo}. The CEO setup corresponds to the case where for each $n$, the optimization over \eqref{eq:distortion_def} at the $l$th encoder is with respect to the entire family of block-length $n$ encoders $\Ecal^{(n)}_l$. The relevant results for the CEO in the quadratic Gaussian case and the binary symmetric case are as follow:


$\bullet$~{\bf quadratic Gaussian CEO problem.}
The rate-region for the Gaussian CEO setup was derived in \cite{prabhakaran2004rate}, while  the optimal rate-allocation is studied in \cite{1321210}.
For the case of  $\ga = \ga_1=\ldots=\ga_L$, the optimal rate-allocation for a given sum rate constraint $\sum R_l\leq R$ is given by \cite[Eq. 10]{1321210}: \begin{equation} \label{eq:ceo_sum_rate}
R(D^\star) = \frac{1}{2} \log^+ \left(\frac{1}{D^\star} \right)-\frac{L}{2} \log^+ \left(
1+ \f 1 {\al^2}  - \f 1 {\al^2 D^\star }
\right).
\end{equation}
Note that the difference in our setup from that of the CEO problem of \cite{berger1996ceo} is that, in our problem formulation, the $l$th encoder is limited to a specific family $\Ecal_l^{(n)}$ of block-length $n$ encoders at rate $R_l$ which have optimal performance with respect to a local distortion measure in terms of the observable process $Y_l^n$.

$\bullet$~{\bf Hamming binary  multiterminal mismatched problem. }
The channel model in \eqref{eq:binary +bitflip} is studied in \cite{he2015lower} where an lower bound for the rate-distortion trade-off for the case of $L=2$  is obtained as
\eas{
R_1 & \geq h ( \rho \star h^{-1}(1-R_2))-h(D_1) ) \\
R_2 & \geq h ( \rho \star h^{-1}(1-R_1))-h(D_2) ) \\
R_1+R_2 & \geq 1 + h(\rho)-h(D_1)-h(D_2)
\label{eq:binary ceo result}
}
for $\rho=p_1 \star p_2$, and $R_1,R_2\leq 1$.

\section{Gaussian source with Gaussian noise and quadratic distortion}
\label{sec:Gaussian source+Gaussian noise+quadratic distortion}

As a first example of the optimal rate allocation in the mismatched multiterminal source coding problem, we consider the model in \eqref{eq:gaussian+quadratic}
for which  the mDRF is shown in  \cite[Prop. 5.2]{kipnis2016mismatch}.

\begin{prop} \label{prop:dist_multi gaussian}
The mDRF in the distributed encoding of a Gaussian source with observation model formulated in \eqref{eq:gaussian+quadratic} is given by
\begin{equation} \label{eq:Gaussian_dist}
D\left( \mathcal E (R_1 \ldots R_L) \right) =
\left( 1 + \sum_{l=1}^L \ga_l \f{1-2^{-2R_l}  } {1+ \ga_l 2^{-2R_l} }  \right)^{-1}.
\end{equation}
\end{prop}

With Prop. \ref{prop:dist_multi gaussian}, the optimal rate allocation under the sum rate constraint of \eqref{eq:sum rate constraint} is obtained in the next theorem.
%
%
\begin{thm}
\label{thm:Optimal rate allocation gaussian}
Let $L_0$ be the number of coefficients $\ga_l$ equal to $\ga_1$, that is  $L_0=  \sum_{i=1}^L 1_\{\ga_i = \ga_1\}$, and define
\ea{
g(\nu,\ga_l) = \f{\ga_l}{\nu}\left(\ga_l+1-\nu + \sqrt{(\ga_l+1)(\ga_l+1-2\nu)}\right).
\label{eq:g_l}
}
When
\ea{
R\geq \f{L_0}{2}\log^+\ga_1,
\label{eq:R_lower_bound}
}
and if  there exists a $\nu^*$ such that
\ea{
\sum_{l=1}^L R_l^*(\nu^*) = R,
\label{eq:kkt_nu}
}
then, the optimal rate allocation for the model in  \eqref{eq:gaussian+quadratic} under the sum rate constraint in \eqref{eq:sum rate constraint}
is attained by the following choice:
\ea{
R_l^*(\nu)=\lcb \p{
\f 12 \log g(\nu^*,\ga_l)   & 0 \leq \ga_l \leq 1,  & 0 \leq \nu^* \leq \f{2 \ga_l}{\ga_l+1} \\
\f 12 \log g(\nu^*,\ga_l)  &  1 \leq \ga_l,   & 0 \leq \nu^* \leq \f{\ga_l+1}{2} \\
0 &  & {\rm otherwise}
} \rnone
\label{eq:water filling}
}
\end{thm}
\begin{IEEEproof}
The proof is provided in App. \ref{pf:Optimal rate allocation gaussian}.
\end{IEEEproof}
\smallskip
Note that \eqref{eq:R_lower_bound} is always satisfied if $\ga_1\leq1$.
However, for $\ga_1>1$, the solution $\nu^*$ may not exist for small $R$.
{Moreover, we see that if $\ga_1>\ga_i>1$ for some $i$, the sum rate constraint \eqref{eq:kkt_nu} may still not have a solution for some values of $R$ due to the discontinuity of $\sum_{l=1}^L R_l^*(\nu^*)$.}
The result in Th. \ref{thm:Optimal rate allocation gaussian} can be interpreted as follows:  we refer to a remote user which transmits at a rate strictly larger than zero as ``active''. Then the conditions in the r.h.s. of \eqref{eq:water filling} decide which encoders are active as a function of the parameter $\nu^*$.
For a given $\nu^*$, while \eqref{eq:water filling} decides whether a user is active or not, \eqref{eq:g_l} specifies the rates of the active users.
The parameter $\nu^*$ is chosen so that the total rate constraint is met with equality when such a solution exists.

\begin{figure}
\centering
\begin{tikzpicture}
\node at (0,0) {\includegraphics[trim=0cm 0cm 0cm 0cm,  ,clip=true,scale=0.36]{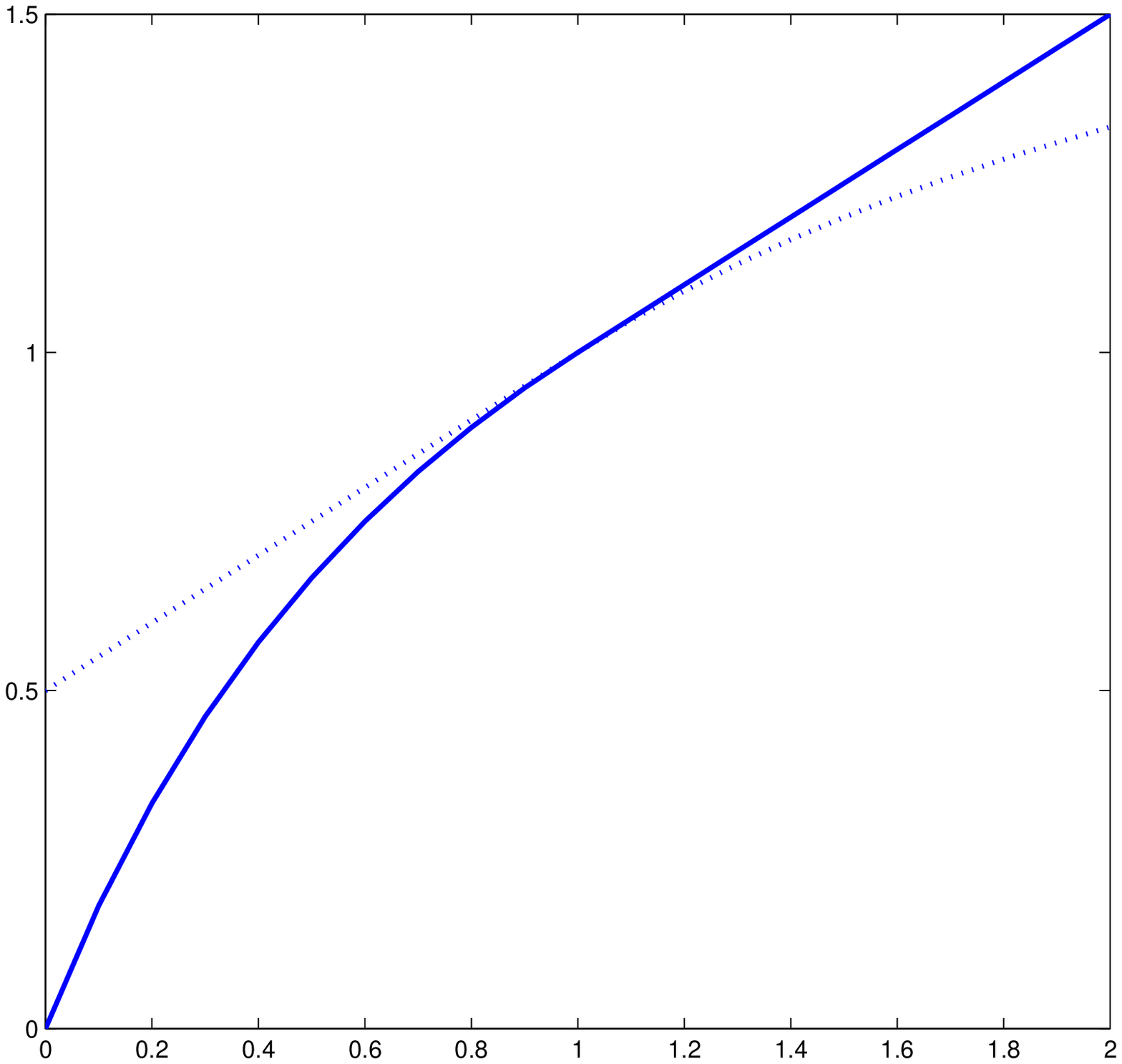}};
\node[rotate=90] at (-3.3,0.3) {$\nu^{*}$ } ;
\node at (0,-3) {\Large$\ga_l$};
\node[rotate = +35] at (+0,+1.5) {\color{blue} Activation threshold};
\node at (+1.5 ,-2.1) {Encoder is active};
\node at (-.9,2.5) {Encoder is not active};
\draw[dashed] (0.15,-1) -- (0.15,1.2);
\node at (-.95,-.9) [text width=1 cm]{\color{blue} inverse region};
\node at (1.2,-.9) [text width=1 cm] {\color{blue} linear region};
%
\end{tikzpicture}
\caption{ The remote user activation threshold in Th. \ref{thm:Optimal rate allocation gaussian}. }
\label{fig:threshold}
\vspace{-.5cm}
\end{figure}

%
An interesting observation is that in \eqref{eq:water filling} a simple threshold decides which encoders are active: for the remote encoders with $0 \leq \ga_l \leq 1$, the activation rule is
$\nu^* \lessgtr 2 \ga_l/(\ga_l+1)$ (inverse in $\ga_l$); while for encoders with $\ga_l>1$, we have the activation rule as  $\nu^* \lessgtr (\ga_l+1)/2$ (linear in $\ga_l$). {The activation rules are illustrated in Fig. \ref{fig:threshold}.}

\begin{figure}
\centering
\begin{tikzpicture}
\node at (0,0) {\includegraphics[trim=0cm 0cm 0cm 0cm,  ,clip=true, width=0.45\textwidth]{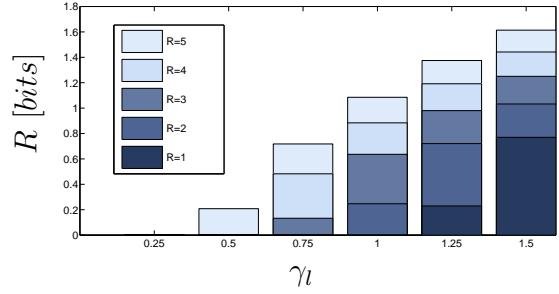}};
\node[rotate=90] at (-3.7,0.3) {\large $R~[bits]$} ;
\node at (0,-2) {\large {$\ga_l$ } };
\end{tikzpicture}
\caption{ 
The optimal rate allocation for the case of $L=5$ and for different values of $\ga_l$ and $R$. 
}
\label{fig:gaussian_alloc}
\vspace{-.5cm}
\end{figure}

The optimal rate allocations for the case of $L=5$ users with various rate budgets are presented in Fig. \ref{fig:gaussian_alloc}. Note that the low-SNR remote encoders are active only for very large rate budgets and are quickly deactivated as the total rate budget decreases.
The result in Fig. \ref{fig:gaussian_alloc} is obtained by noticing that \eqref{eq:g_l} is decreasing in $\nu$, hence the optimal allocation is numerically obtained by setting $\nu=({\ga_1+1})/ 2$ and progressively reducing this value until the sum rate constraint in \eqref{eq:sum rate constraint} is met with equality.

\begin{figure}
\centering
\begin{tikzpicture}
\node at (0,0) {\includegraphics[trim=0cm 0cm 0cm 0cm,  ,clip=true, width=0.34 \textwidth]{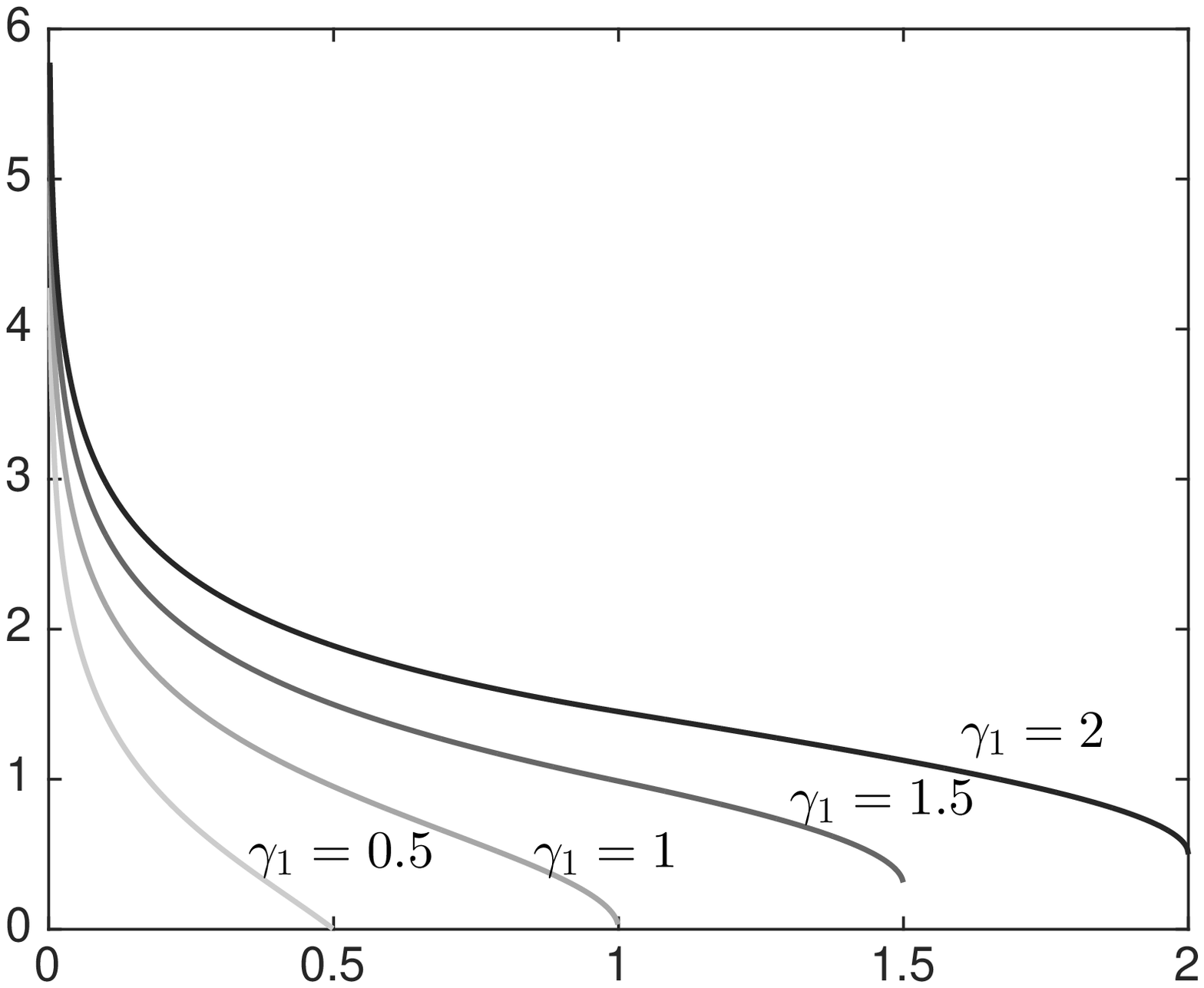}};
\node[rotate=90] at (-3,0.3) {\large $R~[bits]$} ;
\node at (.8,1.7) {Two active encoders};
\node at (0,-2.2) {\large {$\gamma_2$ } };
\end{tikzpicture}
\caption{The total rate budget below which all the rate should be allocated to one user for the case of Gaussian source, $L = 2$, and $\ga_1 \in \{.5,1,1.5,2\}$, $\ga_2 \in [0,\ga_1]$.
}
\label{fig:gaussian_single_active}
\end{figure}

Another scenario we are interested in is the case when the optimal rate allocation only has one active encoder and thus the presence of the second encoder does not further decrease the distortion.
%
For $L = 2$, recall that \eqref{eq:Gaussian_dist} is a function of $R_1$. By minimizing (\ref{eq:Gaussian_dist}), we can then derive an upper bound for the total rate budget below which only one sensor should be active in order to optimally represent the hidden source.
In Fig. \ref{fig:gaussian_single_active} we plot this value of $R$ as a function of $\ga_2$ for various $\ga_1$.
If the total budget is below the plotted line, then the optimal rate allocation is $R_1^*=R$, $R_2^* = 0$.

\section{Binary source with Bit-flipping noise under Hamming distortion}
\label{sec:Binary source+bitflipping noise+Hamming distortion}

Another model of interest for many practical communication settings is the one in which a binary source is observed through Binary Symmetric Channels (BSCs) at the remote sensors and reconstructed at the central unit by minimizing the Hamming distortion.
In order to derive the optimal rate-allocation for the model formulated in \eqref{eq:binary +bitflip}, we first write the mDRF derived in \cite[Prop. 5.3]{kipnis2016mismatch} for the case $\mathbb P(X_i=1) = 1/2$ in the following form:
\begin{prop} \label{prop:dist_multi binary}
The mDRF in the distributed encoding of a binary source with the observation model specified in \eqref{eq:binary +bitflip} is
\ea{
D\left( \mathcal E (R_1 \ldots R_L) \right)
= \Pr \lb \log  F >0  \rb,
\label{eq:bsc_dist}
}
where
\ea{
\log F & = \sum_{l=1}^L  c_l \lb  2 U_l - 1 \rb,
\label{eq:likelihood ration bsc}
%
}
for $c_l =  \log \f {1-q_l} {q_l}$ and where $U_1,\dots,U_L$ are i.i.d binary random variables satisfying
\begin{equation}
\Pr(U_l=1) = q_l =p_l\star D_l,
\label{eq:q}
\end{equation}
where $D_l=h^{-1}(1-R_l)$.
\end{prop}
\begin{IEEEproof}
The proof is a reformulation of the result in \cite[Prop. 3]{kipnis2016mismatch}.  
\end{IEEEproof}

Proposition~\ref{prop:dist_multi binary} provides the following interpretation for the mDRF: each lossy-compressed observation $\Yh_l$ is obtained by passing the observation $X$ through a series of two BSCs: one with cross-over probability $p_l$ and the other with cross-over probability $D_l$, corresponding to a BSC with cross-over probability $q_l$ as defined in \eqref{eq:q}.
The BSC with cross-over probability $D_l$ corresponds to the optimal test-channel in lossy compression of a binary uniform source under Hamming distortion at rate $R_l(D_l)$.
The optimal estimation of $X$ at the central {decoder} is obtained through a log-likelihood ratio test in which each observation is weighted according to a parameter $c_l$ which is controlled by the cross-over probability $q_l$.
Thus, the distortion corresponds to the probability that the log-likelihood ratio fails given the two equally likely observations.

The following theorem describes the optimal rate allocation that minimizes \eqref{eq:bsc_dist}:

\begin{thm}
\label{thm:Optimal rate allocation BSC}
The optimal rate allocation for the model in  \eqref{eq:binary +bitflip} under the sum rate constraint in \eqref{eq:sum rate constraint}
is attained by the following choice:
\ea{
R_l^*=\lcb  \p{
0 &   R<l-1 \\
R-l+1 &  l-1 \leq R < l \\
1 & R\geq l
}
\rnone
\label{eq:Optimal rate allocation BSC}
}
\end{thm}
\begin{IEEEproof}
The result in Th. \ref{thm:Optimal rate allocation BSC} is shown by considering the necessary properties of the optimal solution.
In particular, if a solution is optimal, then there cannot be a remote encoder observing the source through a BSC with cross-over probability $p_{l+1}$ with a positive rate allocated while there also exists an encoder observing the source through a BSC with $p_l<p_{l+1}$
and $R_l<1$.
%
This is because a lower overall distortion would be attained by assigning the rate of user $l+1$ to user $l$ until $R_l=1$.
The remainder of the proof can be found in App. \ref{app:Optimal rate allocation BSC}.
\end{IEEEproof}

As in the result of Th. \ref{thm:Optimal rate allocation gaussian}, the result in Th. \ref{thm:Optimal rate allocation BSC} shows that there exists regimes in which a remote encoder is ignored by the optimal rate allocation if its observation is too noisy or if the available communication rate is too small.

\begin{figure}
\centering
\begin{tikzpicture}
\node at (0,0) {\includegraphics[trim=0cm 0cm 0cm 0cm,  ,clip=true, width=0.36 \textwidth]{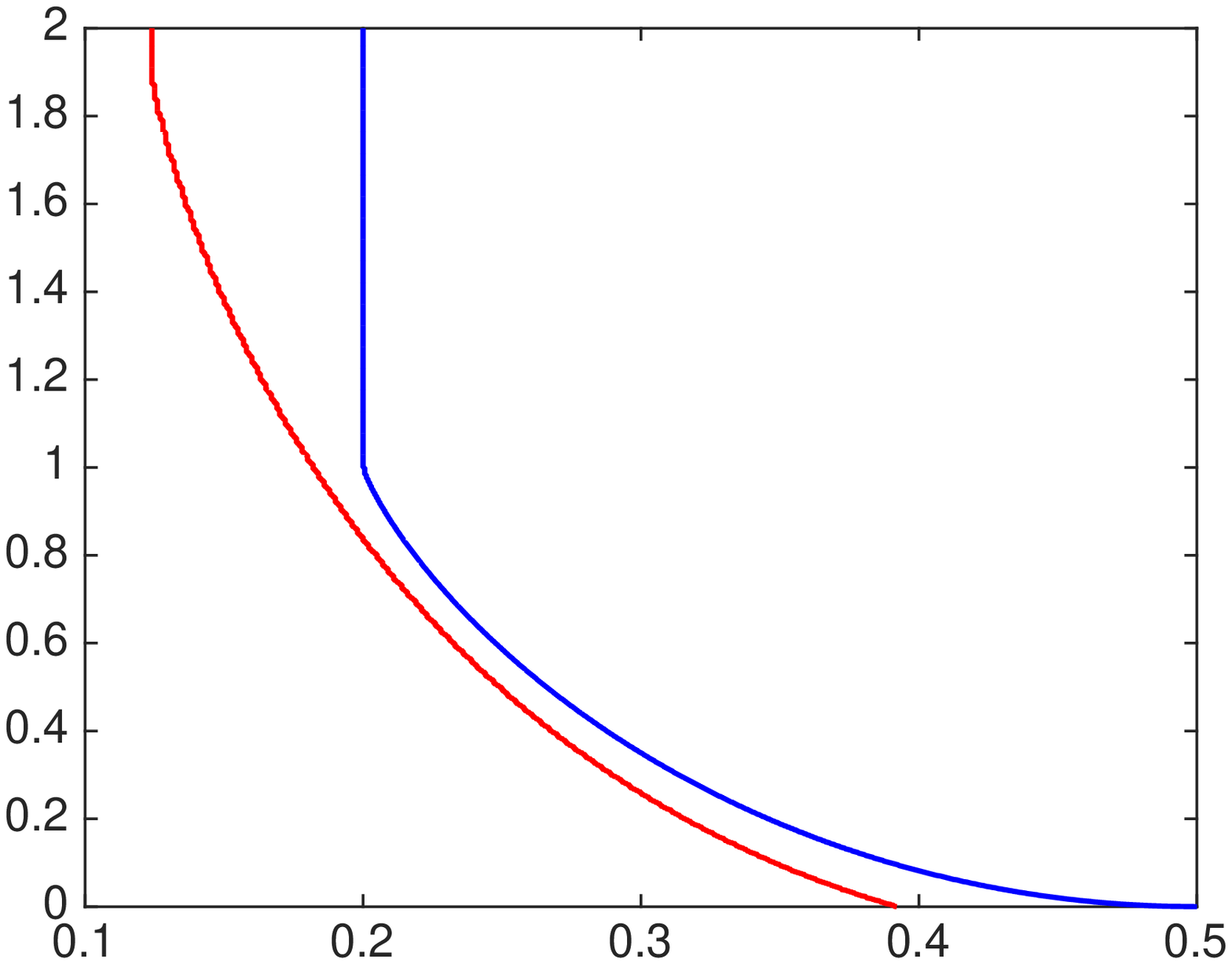}};
\node[rotate=90] at (-3.5,0.3) {\large $R~[bits]$} ;
\node at (0,-2.35) {\large {$D$ } };
\node[rotate=-30, text width=3 cm] at (-0.5,-1.2){\color{red}{CEO  lower bound}};
\node[rotate=-30] at (0,-0.7){\color{blue}{mismatch}};
\end{tikzpicture}
\caption{
The optimal rate allocation for symmetric binary CEO lower bound (red) and the symmetric mismatched remote source coding problem (blue)
for  $L=2$ and $p=0.2$ in Sec. \ref{sec:Binary source+bitflipping noise+Hamming distortion}.
}
\label{fig:binary_comp}
\end{figure}

Note that the rate for the symmetric case $p_1=\ldots=p_l=p$ does not result in the same rate being assigned to all the remote encoders. This is in stark contrast with the optimal solution for the CEO problem: in Fig. \ref{fig:binary_comp} we illustrate the symmetric binary case of \eqref{eq:Optimal rate allocation BSC} and compare it with the CEO result in \eqref{eq:binary ceo result}.

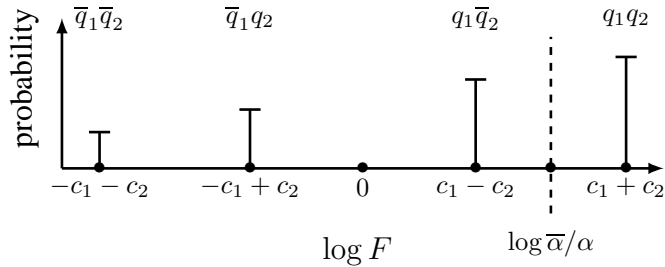
\begin{figure}
\centering
\begin{tikzpicture}[node distance=1 cm,auto,>=latex]

\draw[->, line width=1.0pt] (-4,0) -- (+4,0) ;
\draw[->, line width=1.0pt] (-4,0) -- (-4,2) ;
\node at (0,0){$\bullet$} ;
\node at (0,-.25){$0$} ;
\node at (1.5,0){$\bullet$} ;
\node at (1.5,-.25){$c_1-c_2$} ;
\node at (1.5,2){$q_1\qo_2$} ;
\draw[-|, line width=1.0pt] (1.5,0) -- (1.5,1.2) ;

\node at (-1.5,0){$\bullet$} ;
\node at (-1.5,-.25){$-c_1+c_2$} ;
\node at (-1.5,2){$\qo_1 q_2$} ;
\draw[-|, line width=1.0pt] (-1.5,0) -- (-1.5,.8) ;

\node at (-3.5,0){$\bullet$} ;
\node at (-3.5,-.25){$-c_1-c_2$};
\node at (-3.5,2){$\qo_1\qo_2$};
\draw[-|, line width=1.0pt] (-3.5,0) -- (-3.5,0.5) ;

\node at (+3.5,0){$\bullet$} ;
\node at (+3.5,-.25){$c_1+c_2$} ;
\node at (+3.5,2){$q_1 q_2$} ;
\draw[-|, line width=1.0pt] (+3.5,0) -- (3.5,1.5) ;

\node at (+2.5,0){$\bullet$} ;
\node at (+2.5,-1){$\log \alb / \al$} ;
\node[rotate=90] at (-4.5,1.2) {\large{probability}} ;
\node at (0,-1.1) {\large {$\log F$ } };
\draw[dashed, line width=1.0pt] (+2.5,-.6) -- (2.5,1.75) ;

 \end{tikzpicture}
\caption{An example of $\log F$ for the case in which the source is Bernoulli with parameter $\al$.}
\label{fig:distribution F}
\vspace{-.25 cm}
\end{figure}

Th. \ref{thm:Optimal rate allocation BSC} holds for a symmetric source: if the source follows a general Bernoulli distribution with parameter $\al<1/2$, then the solution in Th. \ref{thm:Optimal rate allocation BSC} is no longer optimal.
This can be clearly shown for the case $L=2$: in this case the distribution of $\log F$ is represented in Fig. \ref{fig:distribution F}. We can see that
the  support of $\log F$ has $4$ points:
$
(-c_1-c_2, \ -c_1+c_2, \  +c_1-c_2, \ +c_1+c_2)
$, with probability
$
(\qo_1 \qo_2, \ \qo_1 q_2, \ q_1 \qo_2, \ q_1 q_2)
$
respectively.

\begin{figure}
\centering
\begin{tikzpicture}
\node at (0,0) {\includegraphics[trim=0cm 0cm 0cm 0cm,  ,clip=true, width=0.36 \textwidth]{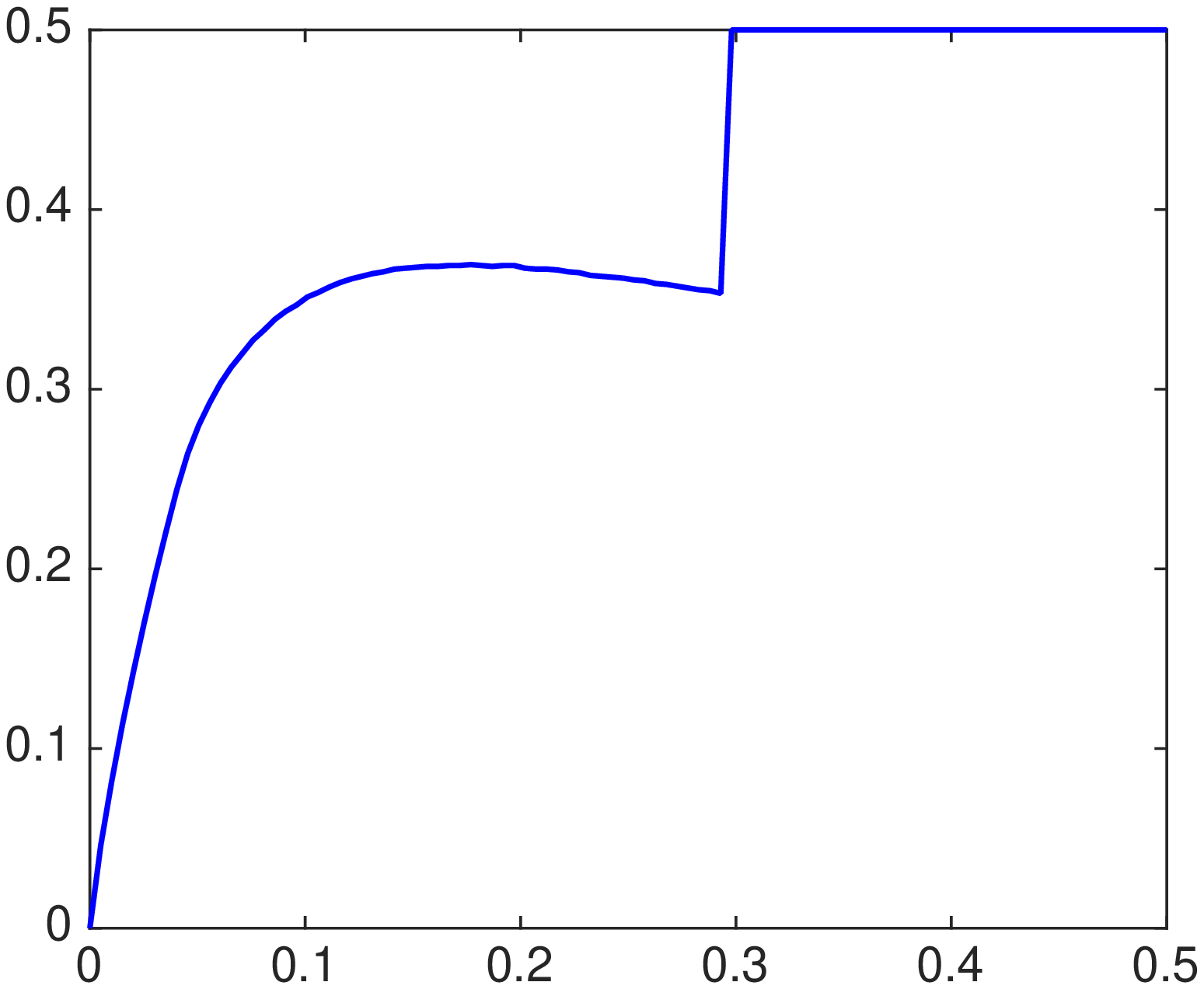}};
\node[rotate=90] at (-3.3,0.3) {\large $R_1~[bits]$} ;
\node at (0,-2.5) {\large {$\al$ } };
\end{tikzpicture}
\caption{
The optimal rate allocation of $R_1$ for a Bernoulli source with parameter $\al \in [0, 1/2]$ and the total sum-rate $R=0.5$ while the cross-over probabilities are $p_1=1/5$ and $p_2=1/3$.
}
\label{eq:bin optimal rates}
\end{figure}

Assuming that $c_1>c_2$, the  log-likelihood threshold $\log \alb / \al$ corresponds to a positive value which can be located either in the interval $[0 , +c_1-c_2]$ or in $[+c_1-c_2,  \ c_1+c_2]$.
In the first interval, the probability of error
is $q_1$, regardless of the value of $q_2$ while, in the second interval, the probability of error is $q_1q_2$.
When minimizing the probability $q_1$, the best rate allocation strategy is to assign  the full rate to $R_1$ while the rate assigned to $R_2$ does not affect the overall distortion.
In general, this choice is not optimal when minimizing $q_1 q_2$.
Since determining the optimal choice for the latter case in closed form is challenging, we show this result numerically in Fig. \ref{eq:bin optimal rates}.

\section{Conclusions}
\label{sec:Conclusions}
We have considered the optimal rate allocation schemes for two mismatch multiterminal source coding problems.
In particular, for an iid Gaussian source observed through AWGN channels under a quadratic distortion measure, we derived the optimal allocation scheme in which remote users are assigned a positive rate according to a threshold that varies with the observation quality.
%
For an iid symmetric binary source observed through binary symmetric channels under Hamming distortion, we also determine the optimal rate allocation and, for this model, the remote encoders with less noisy observations compress their observations at the minimum distortion and thus the number of active encoders equals the number of total bits plus one.
%
%

\bibliographystyle{IEEEtran}
\bibliography{mismatched}
\onecolumn
\appendix
\subsection{Proof of Theorem \ref{thm:Optimal rate allocation gaussian}}\label{pf:Optimal rate allocation gaussian}
Let $\bm R = (R_1, \ldots, R_L)$. The optimal allocation of rate $\bm R^*$ minimizes the minimum mean square error in (\ref{eq:Gaussian_dist}) and
 hence is the solution to the optimization problem which can be formulated as follows:
\begin{equation}\label{opt:R}
\begin{array}{rl}
f_L(\bm R)&\triangleq- \left( 1 + \sum_{l=1}^L \frac{\ga_l(2^{2R_l}-1)}{\ga_l+2^{2R_l}} \right),\\
{\rm subject\ to}& 1^\intercal \bm R \leq R,\\
& {\bm R}\succeq 0.
\end{array}
\end{equation}
The first and second order partial derivatives of the target function $f_L(\bm R)$ can be expanded as
\begin{equation}
\frac{\partial f_L}{\partial R_l} = -{2\ga_l(\ga_l+1)}\frac{4^{R_l}}{(4^{R_l}+\ga_l)^2},
\label{eq:monotonically decreasing}
\end{equation}
\begin{equation}
\frac{\partial^2 f_L}{\partial R_l^2} = 4\ga_l(\ga_l + 1) \frac{4^{R_l}(4^{R_i}-\ga_l)}{(4^{R_l}+\ga_l)^3},
\end{equation}
and
\begin{equation}
\frac{\partial^2 f_L}{\partial R_l \partial R_k} = 0, {\rm for}\ l\neq k.
\end{equation}

The Lagrangian of the optimization problem in (\ref{opt:R}) is
\begin{equation}
L({\bm R},{\bm\lambda},{\nu} ) = f_L(\bm R) + \sum_{l = 1}^{L} \lambda_l(-R_l)+ \nu (\sum_{l = 1}^L R_l - R).
\end{equation}
Let $\bm R^*$ and $(\bm\lambda^*,\nu^*)$ be the primal and dual optimal points with zero duality gap. The KKT conditions can be expanded as follow:
\eas{
\bm R^*&\succeq 0,\\
\sum_{l = 1}^L R_l^* -R &= 0,
\label{eq:rate constraint} \\
\bm\lambda^* &\succeq 0,\\
\nu^* &\geq 0,\\
\lambda^*_l R_l^* &= 0,\ l = 1,\ldots,L,
\label{eq:positive rates}\\
\frac{\partial f_L(R^*) }{\partial R_l} - \lambda_l^* +\nu^* &= 0,\ l = 1,\ldots,L.
\label{eq: sum to R}
}

When a rate $R_l>0$, we must have $\la_l=0$ because of \eqref{eq:positive rates}, in which case \eqref{eq: sum to R} yields

\ea{
\nu^* - {2\ga_l(\ga_l+1)}\frac{4^{R_l^*}}{(4^{R_l^*}+\ga_l)^2}&= 0.
\label{kkt:simplified}
}
The solution of \eqref{kkt:simplified} in $4^{R_l^*}$, if it exists, is
\ea{
4^{R_l^*} =\frac{\ga_l}{\nu^*}\left(\ga_l+1-\nu^*\pm\sqrt{(\ga_l+1)(\ga_l+1-2\nu^*)}\right)\triangleq g_{\pm}(\nu^*,\ga_l)
\label{eq:solution quadratic}
}
for
\ea{
0 \leq \nu^* \leq  \f {\ga_l+1} 2.
}
Note that
\[
\frac{\partial^2 f_L}{\partial R_l^2}\left(\frac{1}{2}\log(g_{+}(\nu^*,\ga_l)\right)>0,
\]
\[
\frac{\partial^2 f_L}{\partial R_l^2}\left(\frac{1}{2}\log(g_{-}(\nu^*,\ga_l)\right)<0,
\]
so that only the $g_{+}$ solution corresponds to the minimal of the target function.
We can  therefore drop the subscript ${}_+$ in \eqref{eq:solution quadratic} and simply refer to $g_{+}(\nu^*,\ga_l)$ as $g(\nu^*,\ga_l)$ as in \eqref{eq:g_l}.

The solution in \eqref{eq:solution quadratic} corresponds to a positive rate only when $g(\nu^*,\ga_l)\geq 1$, which can be equivalently written as
\ea{
\gamma_l \sqrt{\gamma_l+1-2 \nu^*} \geq (\nu^*-\ga_l)\sqrt{\gamma_l+1}.
\label{eq:positive rate}
}
This is always the case when $\ga_l>1$. However, when $\ga_l<1$ it requires the additional condition
\ea{
0  \leq \nu \leq  \f {2 \ga_l}{\ga_l+1}.
}
Since
\ea{
\f {\ga_l+1} 2 \geq \f {2 \ga_l}{\ga_l+1},
}
we have that  the optimal rate allocation is
\ea{
R_l^*(\nu)=\lcb \p{
\f 12 \log g(\nu^*,\ga_l)  & 0 \leq \ga_l \leq 1  & 0 \leq \nu^* \leq \f{2 \ga_l}{\ga_l+1} \\
\f 12 \log g(\nu^*,\ga_l)   &  \ga_l \geq 1  & 0 \leq \nu^* \leq \f{\ga_l+1}{2} \\
0 &  & {\rm otherwise}
} \rnone
\label{eq:optimal appendix}
}
where $\nu^*$ is determined by the following equation:
\begin{equation}\label{eq:waterlevel}
\sum_{l = 1}^L R_l^*(\nu) = R_{\rm T}(\nu)=R.
\end{equation}
Additionally, since $\f{\partial g(\nu,\ga_l)}{\partial \nu}<0$, we also have that
\ea{
g(\nu,\ga_l)\geq g\left(\f{\ga_l+1}{2},\ga_l\right) = \ga_l.
\label{eq:g and ga}
}
Note that \eqref{eq:waterlevel} can be satisfied only for certain values of $R$.
More specifically, \eqref{eq:g and ga} and \eqref{eq:optimal appendix} together imply
\ea{
R_l \geq \f 12 \log (\ga_l),
\label{eq:bound Rl}
}
which corresponds to the conditions under which the problem is convex.
The bound in \eqref{eq:bound Rl} is not tight for $\ga_l<1$ but could possibly be tight for $\ga_l>1$.
Fortunately, we have
\ea{
g(\nu,\ga_l) \geq  \ga_l,  \quad \nu \in \lsb 0,\f{ \ga_1+1}2 \rsb
}
since
\ea{
g(\nu,\ga_l)-\ga_1 = \f{\ga_l}{\nu}\lb \ga_l+1-2 \nu + \sqrt{(\ga_l+1)(\ga_l+1-2\nu)} \rb,
}
which is positive defined for  $\nu \in [0,\f{ \ga_1+1}2 ]$.
Therefore, we will need the sum rate to be lower bounded by $\f{L_0}{2}\log\ga_1$ in order for the sum rate constraint to be met with equality.
It is also necessary to point out that if there exists $i$ such that $\ga_1>\ga_i>1$, then $R_{\rm T}(\nu)$ in \eqref{eq:waterlevel} is a function with more than one discontinuity point.
Hence, it is possible that the sum rate constraint can not be met with equality for some $R$ even if it is larger than the threshold $\f{L_0}{2}\log\ga_1$.

\subsection{Proof of Theorem \ref{thm:Optimal rate allocation BSC}}
\label{app:Optimal rate allocation BSC}

The proof is constructed by contradiction: we assume that in the optimal rate allocation there exists an $l$ such that
\eas{
R_l^* & =R_l < 1 \\
R_{l+1}^* & =R_{l+1}>0,
}{\label{eq:contraddiction}}
and show that such an assignment cannot be optimal.
In particular, we show that there exists an assignment
\eas{
R_l & =\Rt_l=\Rh_l +\ga \\
R_{l+1}& =\Rt_l=\Rh_{l+1} - \ga
}{\label{eq:new assigment}}
for  some small $\ga$ which  provides a lower overall distortion.

From \eqref{eq:bsc_dist} and  \eqref{eq:likelihood ration bsc} we have that $\Pr[\log F>0]=\Pr[Q+K>0]$ for
\ea{
Q =\sum_{j=1, \ j \notin \{l,l+1\} }^L  c_j \lb 2 U_j-1 \rb,
}
and
\ea{
K=c_l \lb 2 U_l-1 \rb + c_{l+1} \lb 2 U_{l+1}-1 \rb.
\label{eq:K}
}

We next show that there exists a choice of $\de_l,\de_{l+1}$ in \eqref{eq:new assigment}  such that using the same $X$ estimate as if $K$ was drawn according to
 \eqref{eq:contraddiction} but for the rate assignment in  \eqref{eq:new assigment} produces a lower distortion.
In other words, we choose a new rate assignment, use a sub-optimal estimator but still obtain a lower distortion, from which it follows that the assignment in
\eqref{eq:contraddiction} cannot be optimal.

\smallskip

Consider first the case $D_l>D_{l+1}$ and $q_{l+1}>q_l$, which is the case when the worse channel is described with a lower distortion than the better channel.
In this case the distribution of $K$ is
\ea{
K=\lcb\p{
-c_l-c_{l+1} & \qo_l \qo_{l+1} \\
-c_l+c_{l+1} & \qo_l q_{l+1} \\
+c_l-c_{l+1} & q_l \qo_{l+1} \\
+c_l+c_{l+1} & q_l q_{l+1}
}\rnone
\label{eq:k distribution}
}
with $-c_l+c_{l+1} \leq 0$ since $q_{l+1}>q_l$ implies $c_l>c_{l+1}$.

The assignment in \eqref{eq:new assigment}  attains the distortions
\eas{
\Dt_l & =  \Dh_l+\de_l \leq \Dh_l+\de_{l+1}
\label{eq:inequality delta 1}\\
\Dt_{l+1} & =  \Dh_{l+1}-\de_{l+1}
}{\label{eq:inequality delta 1 tot}}
for some $\de_l, \de_{l+1}>0$ and where the inequality in \eqref{eq:inequality delta 1}  follows since $D_l>D_{l+1}$ and an increase in rate increases $\Dt_l$ more than a decrease in rate reduces $D_{l+1}$.

Next we want to show that there exists a choice of $\gamma$ corresponding to a $\de_{l+1}$ for which  the probability that $F$ takes the  positive values
$+c_l-c_{l+1}$ and $+c_l+c_{l+1}$ is smaller with the assignment in  \eqref{eq:new assigment} than with the assignment in \eqref{eq:contraddiction}.
If we can find such a $\gamma$, then the assignment is certainly not optimal, since a suboptimal source estimator can produce a lower distortion.
Given the symmetry in the distribution of $K$, this implies determining $\de_{l+1}$ so that $q_l \qo_{l+1}< \qt_l \widetilde{\qo}_{l+1}$ and $q_l q_{l+1} \leq \qt_l \qt_{l+1}$.
With  the assignment in \eqref{eq:inequality delta 1 tot}, we have that $q_l \qo_{l+1}< \qt_l \widetilde{\qo}_{l+1}$, so that the desired result is shown by showing that 
there exists a $\gamma$ in \eqref{eq:new assigment} for which also $q_l q_{l+1} \leq \qt_l \qt_{l+1}$.
This can be done by showing that the derivative of $\lb p_l \star (D_l+\de_{l+1})\rb \lb  p_{l+1} \star (D_{l+1}-\de_{l+1}) \rb$ is decreasing in $\delta_{l+1}$.
Indeed this derivative is equal to
\ea{
& \f {\partial  \lb p_l \star (D_l+\de_{l+1})\rb \lb  p_{l+1} \star (D_{l+1}-\de_{l+1}) \rb} {\partial \de_{l+1}} \\
& =q_l - q_{l+1} -2 (1-2 p_l)(1-2 p_{l+1}) \de_{l+1}-2 p_{l+1} (1-2 p_l) D_l+2 p_l (1-2 p_{l+1})D_{l+1}, \nonumber
}
which is negative when  $q_{l+1}>q_l$,  $p_{l+1}>p_l$ and $D_l>D_{l+1}$ as per assumption.

%

\smallskip

Consider now the case in which $q_{l+1}>q_l$ but $D_l<D_{l+1}$: also for this case  the   better channel is described with a lower distortion but an even lower distortion can be achieved by further reducing the distortion of the better channel.
That is
\eas{
\Dt_l & =  \Dh_l+\de_l  \\
\Dt_{l+1} & =  \Dh_{l+1}-\de_{l+1}\leq \Dh_{l+1}-\de_{l}.
\label{eq:inequality delta 2}
}
%
Next we want to show that there exists a $\de_{l}$ for which $\lb p_l \star (D_l+\de_l)\rb \lb  p_{l+1} \star (D_{l+1}-\de_l) \rb$ is decreasing in $\de_l$.
%
In this case the derivative becomes
\ea{
& \f {\partial \lb p_l \star (D_l+\de_l)\rb \lb  p_{l+1} \star (D_{l+1}-\de_l) \rb}{ \partial \de_1 }  \\
& = q_l - q_{l+1}+2 p_l(p_l-p_{l+1})  -2 (1-2 p_l)(1-2 p_{l+1}) \de_{l+1}-2 (p_{l+1}-p_l) (1-2 p_l) D_l, \nonumber
}
which is again negative {defined} when $q_{l+1}>q_l$ and $p_l>p_{l+1}$ as assumed.
%
%

In the derivations above, we have shown that, regardless of the assignment in \eqref{eq:contraddiction}, one would obtain the choice policy as stated in the main theorem.

\end{document}